\documentclass[a4paper]{jpconf}
\usepackage{graphicx}
\DeclareMathAlphabet{\mathcal}{OMS}{cmsy}{m}{n}
\usepackage{amsmath}
\DeclareMathAlphabet\mathrsfso{U}{rsfso}{m}{n}
\begin{document}
\title{Measuring the skewness dependency of Generalized Parton Distributions}

\author{S Zhao$^1$ and E Voutier$^{1,2}$}

\address{$^{1}$Institut de Physique Nucl\'{e}aire d'Orsay, CNRS-IN2P3, Universit\'{e} Paris-Sud \& Paris-Saclay, 91406 Orsay, France}

\ead{$^1$zhao@ipno.in2p3.fr, $^2$voutier@ipno.in2p3.fr}

\begin{abstract}
Generalized Parton Distributions (GPDs) have emerged over the 1990s as a powerful concept and tool to study nucleon structure. They provide nucleon tomography from the correlation between transverse position and longitudinal momentum of partons. The Double Deeply Virtual Compton Scattering (DDVCS) process consists of the Deeply Virtual Compton Scattering (DVCS) process with a virtual photon in the final state eventually generating a lepton pair, which can be either an electron-positron or a muon-antimuon pair. The virtuality of the final time-like photon can be measured and varied, thus providing an extra lever arm and allowing one to measure the GPDs for the initial and transferred momentum dependences independently. This unique feature of DDVCS is of relevance, among others, for the determination of the distribution of nuclear forces which is accessed through the skewness dependency of GPDs. This proceeding discusses the feasibility and merits of a DDVCS experiment in the context of JLab 12 GeV based on model-predicted pseudo-data, and the capability of extraction of Compton Form Factors based on a fitter algorithm.

\end{abstract}

\section{Introduction}

With the emergence of the Generalized Parton Distributions (GPDs) formalism \cite{GPD1,GPD2,JiSum,JiSum2} and its associated experimental program, a significant progress has been made in the research of nucleon structure since two decades ago. The GPDs are the structure functions of the nucleon accessed mainly in the deeply exclusive leptoproduction of a photon or a meson, and parametrizing the complex non-perturbative QCD partonic dynamics and structure of the nucleon. They provide nucleon tomography from the correlation between transverse position and longitudinal momentum of partons \cite{Tomo}. As a result of these position-momentum correlations, GPDs provide a way to measure the unknown orbital momentum contribution of quarks to the total spin of the nucleon through Ji's sum rule \cite{JiSum}. They also enable indirect access to one of the gravitational form factors encoding the shear forces and pressure distribution on the quarks in the proton \cite{Nature}.

\begin{figure}[b]
\centering
\begin{minipage}{.45\textwidth}
\centering
\includegraphics[width=0.85\textwidth]{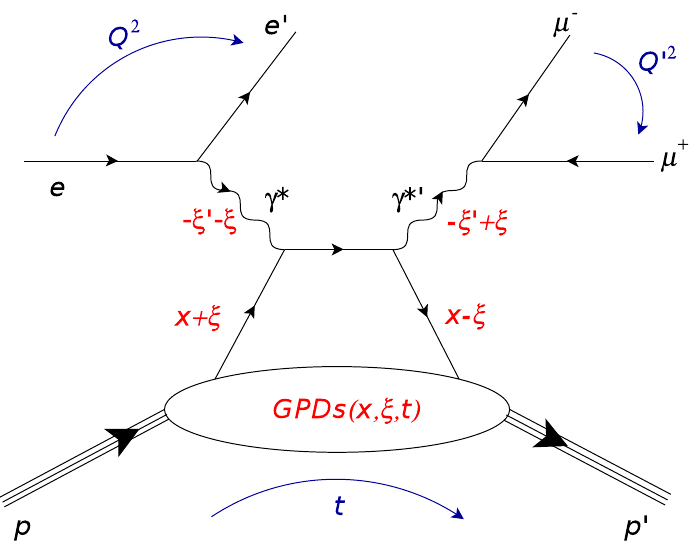} 
\caption{\label{fig1_1}The handbag diagram symbolizing the DDVCS direct term with di-muon final states (there is also a crossed diagram where the final photon is emitted from the initial quark).}
\end{minipage}
\hspace{1cm}
\begin{minipage}{.45\textwidth}
\includegraphics[width=0.49\textwidth]{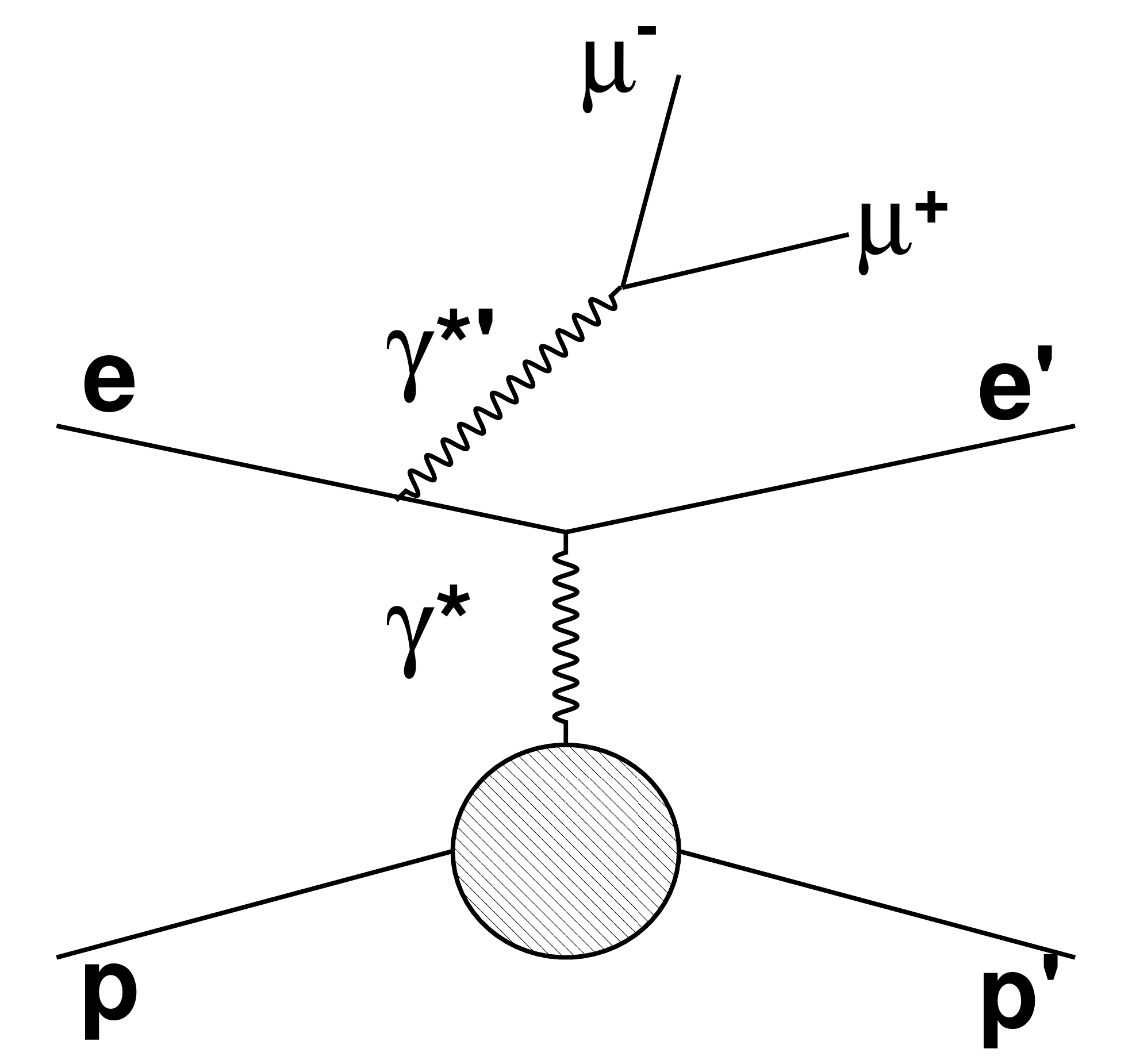} 
\includegraphics[width=0.49\textwidth]{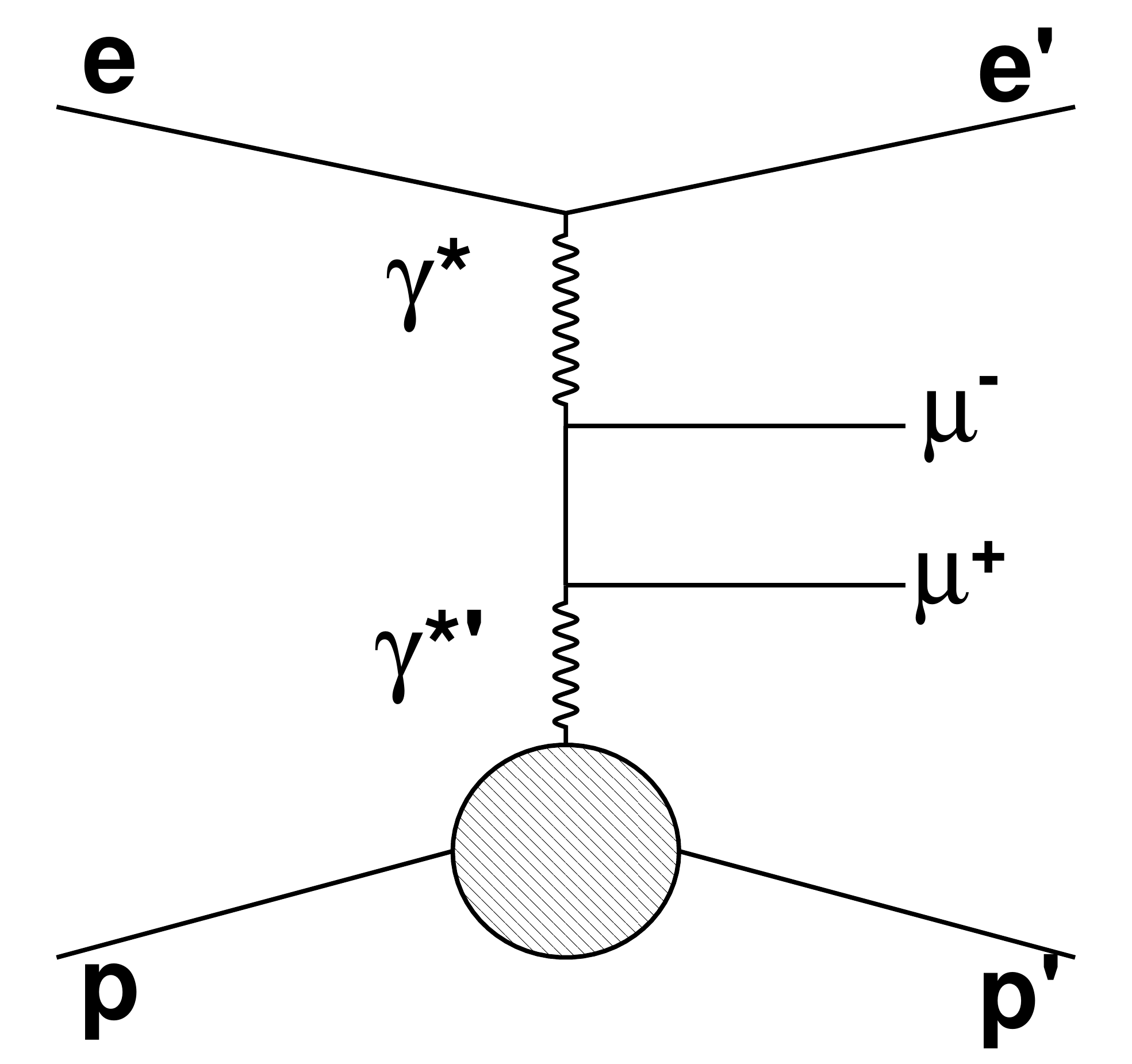} 
\caption{\label{fig1_2}Left: the BH1 process (there is also the process where the time-like photon is emitted from the scattered electron). Right: the BH2 process (there is also the process where the muon-pair exchange their charge).}
\end{minipage}
\end{figure}

There are essentially three experimental golden channels for direct measurements of GPDs: the electroproduction of a photon $eN\rightarrow eN\gamma$ which is sensitive to the deeply virtual Compton scattering (DVCS) amplitude, the photoproduction of a lepton pair $\gamma N\rightarrow l\bar{l}N$ which is sensitive to the timelike Compton scattering (TCS) amplitude, and the electroproduction of a lepton pair $eN\rightarrow eNl\bar{l}$ which is sensitive to the double deeply virtual Compton scattering (DDVCS) amplitude. Only the latter provides the framework necessary for an uncorrelated measurement of a GPD($x,\xi,t$) as a function of both scaling variable $x$ and $\xi$ \cite{DDVCS1,DDVCS2,DDVCS3}. They are the average and the transferred quark momentum fractions, $\xi$ being also referred as the GPD skewness. The former two reactions cannot entirely serve the purpose of testing the angular momentum sum rule due to the real nature of the final- or initial-state photons, which leads to the restriction $x=\pm \xi$. For instance, the Compton form factor (CFF) $\mathcal{H}$ associated with the GPD $H$ and accessible in DVCS cross section or beam spin asymmetry experiments can be written
\begin{eqnarray}
\relax\mathcal{H}(\xi,\xi,t)=\sum_{q}e_q^2&\bigg\{&\mathcal{P}\int_{-1}^1dx~H^q(x,\xi,t)\bigg[\frac{1}{x-\xi}+\frac{1}{x+\xi}\bigg]
\nonumber\\
&&-i\pi\big[H^q(\xi,\xi,t)-H^q(-\xi,\xi,t)\big]\bigg\}
\label{eq1}
\end{eqnarray}
where the sum runs over all parton flavors with elementary electrical charge $e_q$, and $\mathcal{P}$ indicates the Cauchy principal value of the integral. While the imaginary part of the CFF accesses the GPD values at $x=\pm \xi$, it is clear from Eq.~\ref{eq1} that the real part of the CFF is a more complex quantity involving the convolution of parton propagators and the GPD values out of the diagonals that is a domain that cannot be resolved unambiguously with DVCS experiments. Because of the virtuality of final state photons, DDVCS provides a way to circumvent the DVCS limitation, allowing to vary independently $x$ and $\xi$. Considering the same GPD $H$, the corresponding CFF for DDVCS process writes
\begin{eqnarray}
\mathcal{H}(\xi',\xi,t)=\sum_{q}e_q^2&\bigg\{&\mathcal{P}\int_{-1}^1dx~H^q(x,\xi,t)\bigg[\frac{1}{x-\xi'}+\frac{1}{x+\xi'}\bigg]
\nonumber\\
&&-i\pi\big[H^q(\xi',\xi,t)-H^q(-\xi',\xi,t)\big]\bigg\}
\label{eq2}
\end{eqnarray}
providing access to the scaling variable $x =\pm \xi'$ ($\xi' \neq \xi$).

The DDVCS process is most challenging from the experimental point of view due to the small magnitude of the cross section and requires high luminosity and full exclusivity of the final state. Moreover, the difficult theoretical interpretation of electron-induced lepton pair production when detecting the $e^+ e^-$ pairs from the decay of the final virtual photon, hampers any reliable experimental study. Taking advantage of the energy upgrade of the CEBAF accelerator, it is proposed to investigate the electroproduction of $\mu^+ \mu^-$ di-muon pairs and measure the beam spin asymmetry of the exclusive $ep\rightarrow e'p'\gamma^* \rightarrow e'p'\mu^+\mu^-$ reaction in the hard scattering regime \cite{Proposal1,Proposal2,Anikin,ZHAO}.

At sufficiently high virtuality of the initial space-like photon and small enough four-momentum transfer to the nucleon with respect to the photon virtuality ($-t \ll Q^2$), DDVCS can be seen as the absorption of a space-like photon by a parton of the nucleon, followed by the quasi-instantaneous emission of a time-like photon by the same parton, which finally decays into a di-muon pair (figure~\ref{fig1_1}). $Q^2$ and $Q'^2$ represent the virtuality of the incoming space-like and outgoing time-like photons respectively. The scaling variable $\xi'$ and $\xi$ write
\begin{eqnarray}
\xi' = \frac{Q^2-Q'^2+t/2}{2Q^2/x_\text{B}-Q^2-Q'^2+t}~~\text{and}~~
\xi  = \frac{Q^2+Q'^2}{2Q^2/x_\text{B}-Q^2-Q'^2+t}
\label{xipxi}
\end{eqnarray}
from which one obtains
$\xi' = \xi  \frac{Q^2-Q'^2+t/2}{Q^2+Q'^2} .$
This relation indicates that $\xi'$, and consequently the CFF imaginary part, is changing sign about $Q^2=Q'^2$, which procures a strong testing ground of the universality of the GPD formalism. 

A further complexity in studying GPDs via DDVCS is that there is an additional significant mechanism contributing to the same final states, the Bethe-Heitler (BH) processes, as shown in figure \ref{fig1_2}. In the BH1 process the time-like photon is radiated by the incoming or scattered electron, and in the BH2 process it is produced within the nuclear field. The BH and DDVCS mechanisms interfere at the amplitude level. However, the BH amplitudes are precisely calculable theoretically at small momentum transfers $t$ considered in this work. 

In this proceeding, the feasibility of a DDVCS experiment at JLab 12 GeV is discussed. Section \ref{sec2} reports model-predicted experimental projections at a certain luminosity with ideal detectors. In section \ref{sec3}, we apply a fitting method to extract the GPD information from the pseudo-data. Preliminary conclusions about this study are drawn in the last section.

\section{Experiment projections}
\label{sec2}

\begin{figure}[b]
\centering
\includegraphics[width=.49\textwidth]{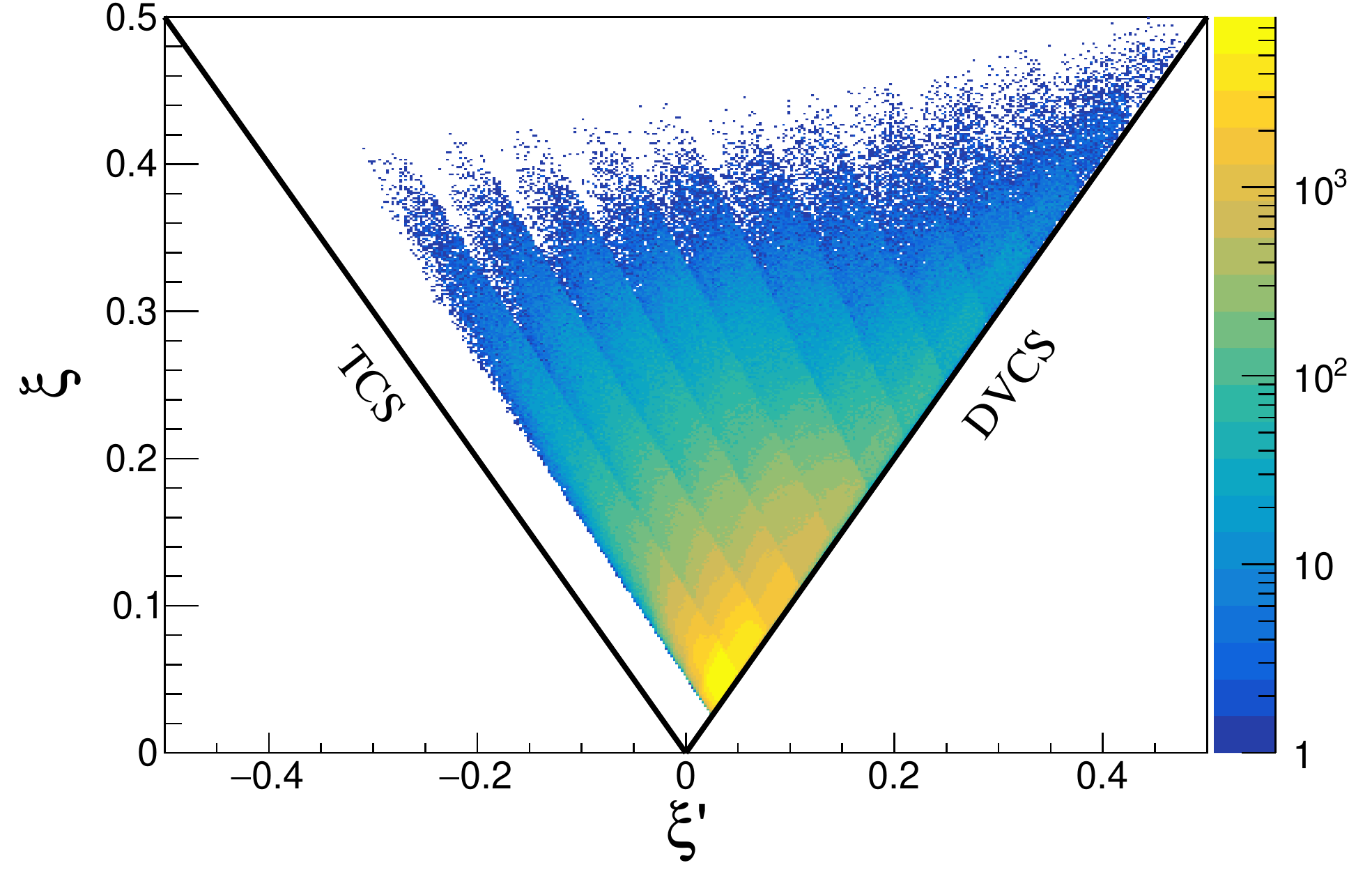} 
\includegraphics[width=.49\textwidth]{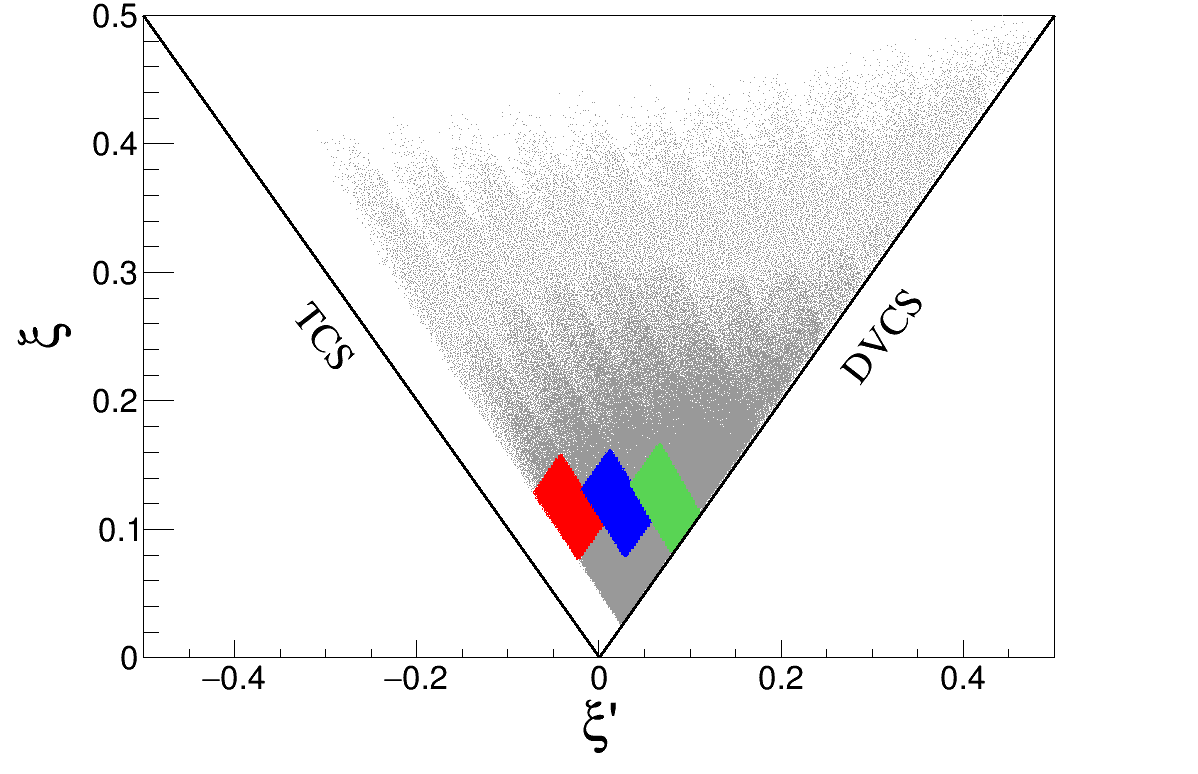} \\
(a)\hspace{7.4cm}(b)
\caption{(a) The distribution of DDVCS count number on ($\xi',\xi$) plane; (b) three example of bins having fixed mean value in $\xi$ and different values in $\xi'$.}
\label{fig2_2}
\end{figure}

A DDVCS event generator based on VGG model \cite{VGG1,VGG2,VGG3} at leading-twist has been developed in order to predict the experimental observables at a certain luminosity for a fixed data taking time. More details about the kinematics and experimental observables we are considering are discussed in \cite{ZHAO}. The improvement we have made in this work is covering the whole kinematic phase space of interests. Figure \ref{fig2_2}(a) depicts the count number distribution on the ($\xi',\xi$) plane, where the solid lines indicate the DVCS correlation ($\xi'=\xi$) and the TCS correlation ($\xi'=-\xi$). All the DDVCS events are within the $|\xi'|<\xi$ region and provide the bins for $\xi$ or $\xi'$ dependency. For instance, three bins in figure \ref{fig2_2}(b) have approximately the same average $\xi$ equalling to 0.11 and different average $\xi'$ being $-0.026$ (red region), $0.022$ (blue region) and $0.087$ (green region), respectively. In the following, the experiment projections and the extraction of CFFs are discussed with respect to the three bins.

\begin{figure}[b]
\centering
\includegraphics[width=.8\textwidth]{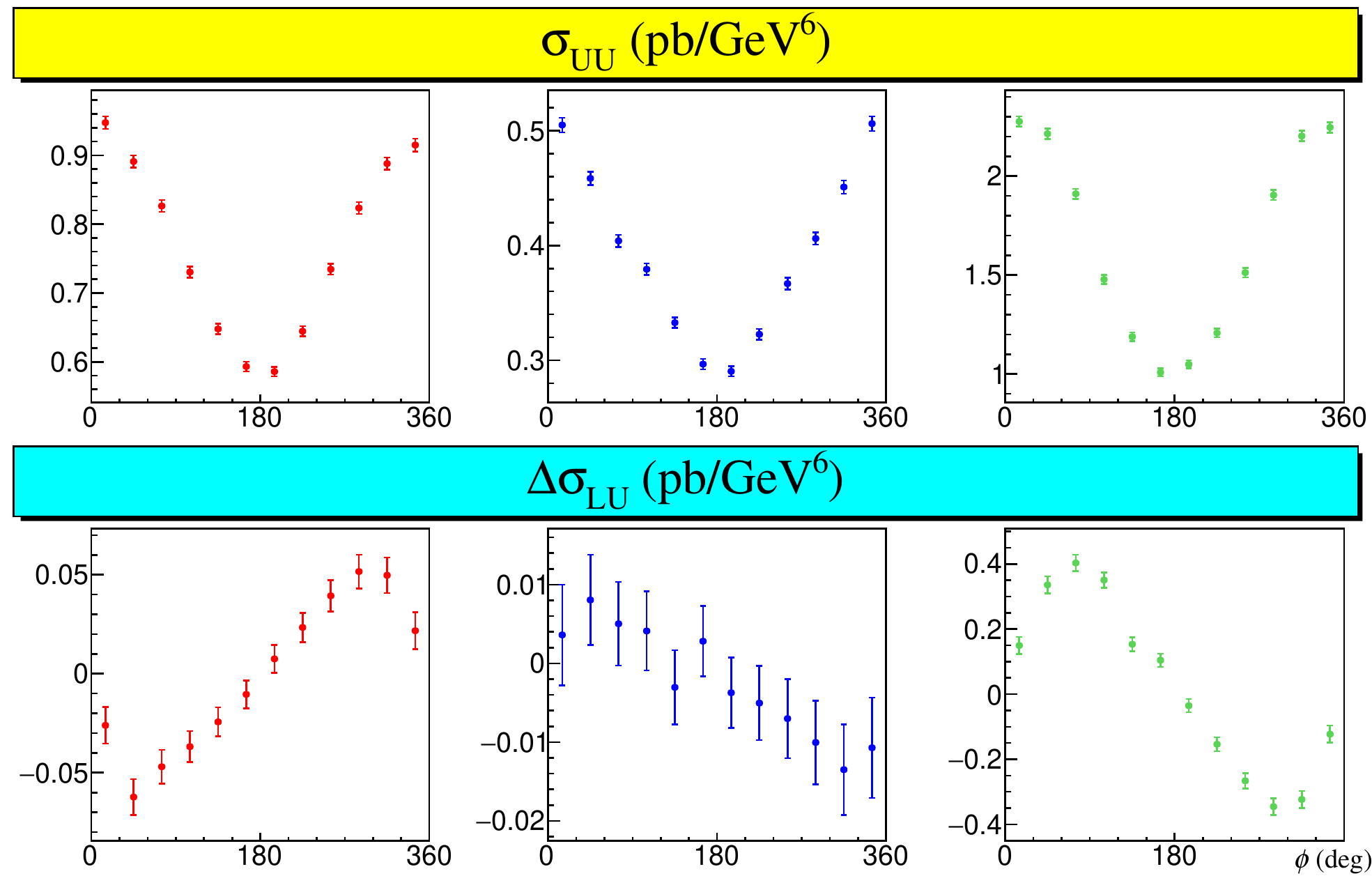} 
\caption{Unpolarized cross section $\sigma_\text{UU}$ (upper panel) and beam spin cross section difference $\Delta\sigma_\text{LU}$ (lower panel) for the three bins in figure \ref{fig2_2}(b). From left to right, they correspond the average $\xi'$ equalling to $-0.026$ (red points), $0.022$ (blue points) and $0.087$ (green points).}
\label{fig2_3}
\end{figure}

The projections have been performed in the ideal situation that all the particles of the final state can be detected with 100\% efficiency. The count-rate calculation was done for a luminosity $\mathrsfso{L} = 10^{36} \text{cm}^{-2}\cdot\text{s}^{-1}$ considering 50 days running time equally distributed between each electron beam polarization. Only unpolarized cross section $\sigma_\text{UU}$ and beam spin cross section difference $\Delta\sigma_\text{LU}$ are discussed in this work. Figure \ref{fig2_3} shows them for the three bins with statistic errors as a function of $\phi$. The central value is smeared according to a Gaussian probability distribution whose standard deviation is equal to the error bar. It indicates that it is possible to obtain DDVCS experimental observables with good precision. Besides, $\Delta\sigma_\text{LU}$ shows the sign change behavior as $\xi'$ increases due to the antisymmetric property of GPD.

\begin{figure}[ht]
\centering
\includegraphics[width=.8\textwidth]{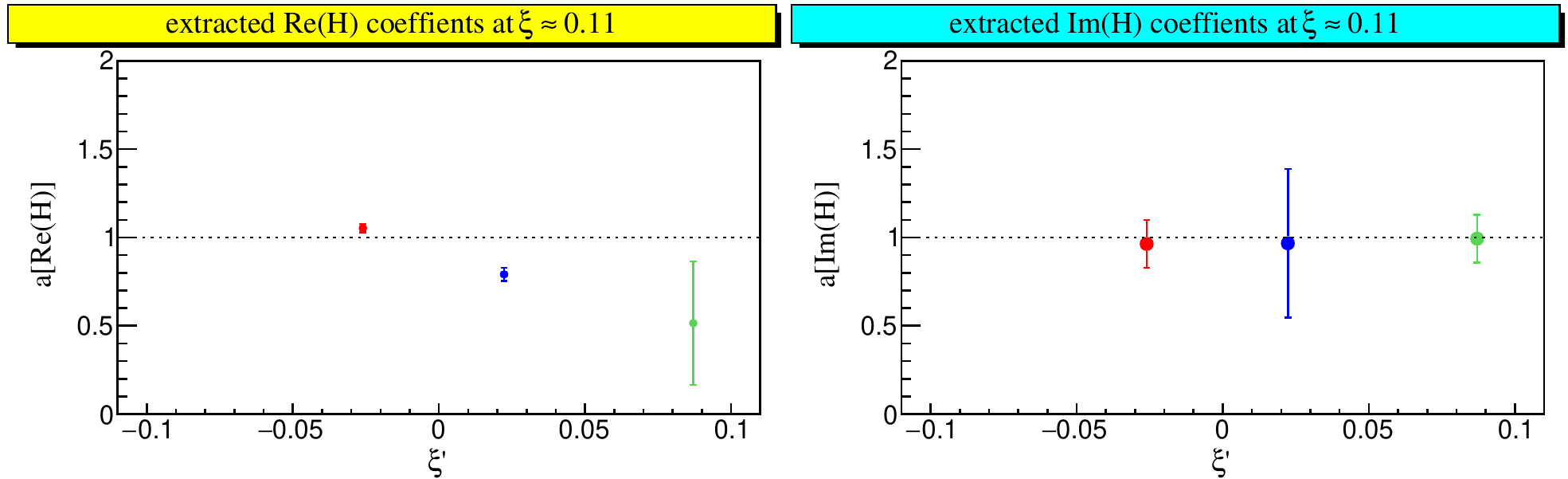} 
\caption{The extracted CFF coefficients for the three bins in figure \ref{fig2_2}(b). The left panel is the real part of CFF $\mathcal{H}$ while the right one is the imaginary part.}
\label{fig3_1}
\end{figure}

\section{CFF extraction}
\label{sec3}

The fitting program is inspired from the one for DVCS process \cite{Fit1}, corresponding to a quasi-model-independent way to extract CFFs. It consists in taking the eight CFFs as free parameters and knowing the well-established BH and DDVCS leading-twist amplitudes to fit, at a fixed kinematics, simultaneously the $\phi$-distributions of several experimental observables. It has been proved in \cite{Fit1} that the fitting program is reliable and powerful to extract all CFFs, given enough observables. If only $\sigma_\text{UU}$ and $\Delta\sigma_\text{LU}$ are available, only CFF $\mathcal{H}$ can be well extracted, with $\sigma_\text{UU}$ being particularly sensitive to the real part and $\Delta\sigma_\text{LU}$ dominated by the imaginary part. Figure \ref{fig3_1} shows the extracted CFF coefficient, defined as the fitted CFF value normalized by the one of the event generator, of the three bins. The dashed line indicates the coefficient being 1, which leads to the fact that the CFF is perfectly recovered. The imaginary part looks obviously well recovered, which means that we are able to extract the imaginary part of $\mathcal{H}$. With respect to the real part, there is a relatively large discrepancy between the fitted and the generated CFF, due to the complex component of $\sigma_\text{UU}$ from different contributions. In order to perform a better extraction of the real part, the beam charge cross section is required since it involves only the interference contribution, which is linear in the CFFs \cite{DDVCS3,ZHAO}.

\begin{figure}[b]
\centering
\includegraphics[width=.4\textwidth]{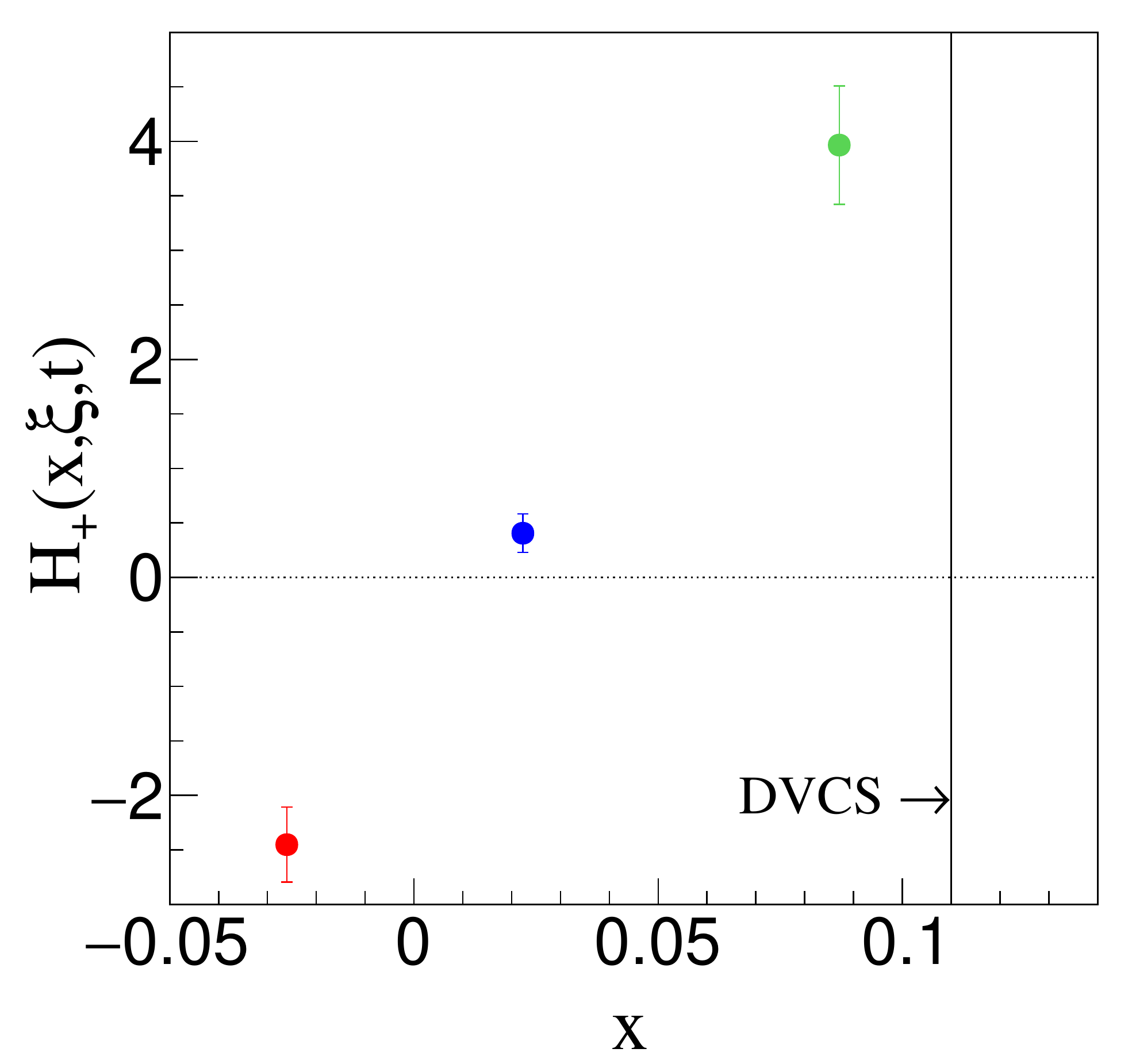} 
\hspace{2pc}%
\begin{minipage}[b]{16pc}\caption{\label{fig_singlet}The proton GPD singlet combination of $H$ with respect to the three bins in figure \ref{fig2_2}(b).}
\end{minipage}
\end{figure}

Eventually, the proton GPD singlet combination of $H$, writing
\begin{eqnarray}
H_+(x,\xi,t)=H(x,\xi,t)-H(-x,\xi,t),
\label{singlet}
\end{eqnarray}
can be extrapolated from the imaginary part of CFF $\mathcal{H}$. In this work $H_+(x,\xi,t)$ at fixed $\xi$ value for three different $x$ is obtained, as shown in figure \ref{fig_singlet}. The vertical solid line stands for the location accessed by DVCS and the three points extracted from the DDVCS pseudo-data circumvent the DVCS limitation and give the decoupled singlet GPD value.

Ji's sum rule \cite{JiSum},
\begin{eqnarray}
\int_{-1}^{1}x\left[H(x,\xi,0)+E(x,\xi,0)\right]dx=\frac{1}{2}\Delta\Sigma+L=J,
\end{eqnarray}
allows access to the contribution of the orbital
momentum of quarks to the spin of the proton. The first Mellin moment of the $H$ GPD can be written \cite{JiSum2}
\begin{eqnarray}
\int_{-1}^{1}xH(x,\xi,t)dx=M_2(t)+\frac{4}{5}\xi^2 d_1(t)
\end{eqnarray}
where $d_1(t)$ encodes the internal shear forces acting on the quarks and their pressure distributions in the proton. Our study shows the feasibility of the access to the GPD $H$ information along the path of integration over $x$ at fixed $\xi$. Therefore, measuring the skewness dependence of GPDs via DDVCS is providing a model-independent access to the strong force distribution and spin origin inside the nucleon.

\section{Conclusion}
\label{sec4}
The model-predicted projections of a DDVCS experiment indicate a high degree of feasibility at a relative challenging luminosity with exclusive final states completely detected. Covering the whole kinematics phase space of interests, we are able to obtain the skewness dependency of GPD $H$ singlet. Applying a further binning approach, we also enable access to the $t$ and $Q^2$ dependency simultaneously. The fitting program is confronted by a severely underconstrained problem and time consuming difficulty, and further work is still ongoing. Beam charge cross section difference is necessary for the extraction of the real part of CFFs.

\section*{Acknowledgement}
This work is supported by the China Scholarship Council (CSC) and the French Centre National de la Recherche Scientifique (CNRS). We would like to express our appreciation to M. Guidal, S. Niccolai and M. Vanderheaghen for helpful discussions.

\end{document}